# Fano-resonance boosted cascaded field enhancement in a plasmonic nanoparticle-in-cavity nanoantenna array and its SERS application

Zhendong Zhu[1,2], Benfeng Bai[1,*], Oubo You[1], Qunqing Li[2], and Shoushan Fan[2]

[1] State Key Laboratory of Precision Measurement Technology and Instruments, Department of Precision Instrument, Tsinghua University, Beijing 100084, China

[2] Tsinghua-Foxconn Nanotechnology Research Center, Department of Physics, Tsinghua University, Beijing 100084, China

\* Correspondence: Dr. Benfeng Bai,

    Department of Precision Instrument, Tsinghua University, Beijing 100084, China

    E-mail: baibenfeng@mail.tsinghua.edu.cn

**Abstract**

Cascaded optical field enhancement (CFE) can be realized in some specially designed multiscale plasmonic nanostructures, where the generation of extremely strong field at nanoscale volume is crucial for many applications, for example, surface enhanced Raman spectroscopy (SERS). Here, we propose a strategy of realizing a high-quality plasmonic nanoparticle-in-cavity (PIC) nanoantenna array, where strong coupling between a nanoparticle dark mode with a high order nanocavity bright mode can produce Fano resonance at a target wavelength. The Fano resonance can effectively boost the CFE in the PIC, with a field enhancement factor up to $5 \times 10^2$. A cost-effective and reliable nanofabrication method is developed with room temperature nanoimprinting lithography to manufacture high-quality PIC arrays. This technique guarantees the generation of only one gold nanoparticle at the bottom of each nanocavity, which is crucial for the generation of the expected CFE. As a demonstration of the performance and application of the PIC array, it is used as active SERS substrate for detecting 4-aminothiophenol molecules. The SERS enhancement factor up to $2 \times 10^7$ is obtained experimentally, verifying the field enhancement and potential of this device.

## INTRODUCTION

Metallic nanostructures supporting surface plasmons (SPs) have attracted much attention due to their peculiar optical properties and important applications.[1,2] SPs can be considered as the resonant photon-induced collective oscillations of free electrons confined on the surfaces of metallic nanostructures, which can be driven by the incident light field. The importance of such photon-electron interactions lies in its ability to concentrate light energy in nanoscale volumes in the metallic nanostructures and thereafter boost the intensity of the optical near field by several orders of magnitude.[2-5] Owing to these properties, the localized SPs can facilitate many applications based on the enhanced light-matter interaction, such as the detection of trace element chemical varieties in the vicinity of metallic nanostructures by surface-enhanced Raman spectroscopy (SERS). The detection limit can be lowered down even to single molecular level.[6,7]

The SP enhanced optical near field (also called the field "hot spot"[6,7]) is typically and predominantly localized in some very small geometric features of metallic nanostructures, such as nano-grooves,[2] nanogaps,[8] and nanotips.[9] Although some advances have been achieved on the fabrication of such nanostructures with extremely small feature sizes,[10] these tiny features are still formidable tasks for nanofabrication. Alternatively, one may realize the extreme field localization and enhancement via a so-called cascaded field enhancement (CFE) mechanism. This concept was probably first proposed by Stockman *et al.* with a self-similar chain of gold nanoparticles (NPs),[2,3] where the light energy can be focused to nanoscale volume via the intercoupling of localized SPs among the aligned NPs with decreasing sizes. The field enhancement is first produced by a relatively large metallic NP that is resonantly excited by the incident light. Then the enhanced field simultaneously excites the field of the adjacent smaller NP via SP coupling. The process continues until the smallest NP is excited so that the light energy is predominantly confined in the tiniest nanogap.[3] Although the self-similar chains of NPs have shown great potential, it is quite challenging to fabricate such nanofocusing systems in practice. Actually, the cascaded energy focusing effect exists not only in the self-similar chains of metal NPs,[3] but also in many other metallic nanostructures with multiscale features, such as the V-grooves,[2] nanostars,[8] ellipsoidal NP dimers,[11] and asymmetric π-shaped metamaterials.[12] So far, only a few experimental verifications of the CFE effect in some geometries were reported.[13] Recently, Kravets *et al.* reported the CFE in a composite plasmonic system where each unit is composed of two or three nanodisks.[14,15] Novotny *et al.* demonstrated the CFE in arrays of gold NPs assembled into dimmer or trimer nanoantennas.[16] Such novel CFE plasmonic nanostructures show great potential for applications such as SERS with single molecular level,[7] high-sensitivity reflective index sensing,[17] and super-resolution imaging.[9,18] In our previous work on a M-shaped nanograting, strong CFE can be achieved by the intercoupling between the localized SPs in two adjacent V-grooves, leading to pronounced enhancement of the SERS sensitivity.[19]

Another effective way of tuning the resonance in a complex plasmonic system is through mode hybridization, which results from the interference between two or more plasmonic modes.[1,5] A plasmonic mode can be either sub-radiant (called the dark mode) or super-radiant (called the bright mode), depending on how strong the incident light can be coupled to the plasmonic mode.[20,21] Analogous to classical Fano resonance system,[1,5,22] plasmonic Fano resonance may also arise from the interference between a continuum of incident photons or a spectrally broad resonance (caused, e.g., by a super-radiant bright mode) with a narrow resonance (caused, e.g., by a sub-radiant dark mode). The intercoupling between the bright mode and dark mode occurs when the near field of one mode overlaps with and excites that of the other mode, which can be generated in some coupled plasmonic nanostructures such as the symmetry breaking nanocavity-NP system,[21] the non-concentric ring-disk cavity,[23] plasmonic nanoshell,[24] NP oligomer chains,[25,] nanovoids,[26-30] nanostars,[31] and nanogratings,[32] nanowell,[44,45] nano-bricks[46,47] The plasmonic Fano interference often results in an asymmetric resonant peak dressing within a broad spectrum.[33]

Complex plasmonic nanostructures where individual tiny metal NPs are coupled to some extended metallic structures are ideal systems for generating both the CFE and mode hybridization, such as the nanoparticle-in-cavity (PIC) nanoantenna array.[26-30] Cole *et al*. have systematically analyzed the excited fields and energies of various localized modes in empty truncated plasmonic nanocavities embedded in a metal film, where various bonding and anti-bonding hybridized modes can be generated.[26-29] Huang *et al*. have theoretically studied a PIC structure and revealed that the CFE can be effectively induced by the intercoupling between the $^0$D mode of the empty void and a "NP mode" induced by the gold NP located at the bottom of the cavity under *oblique incidence*.[30] To verify the CFE effect, they fabricated some PIC structures (with 20 nm Ag NPs in 600 nm Au voids) and used them as SERS substrates to probe 4-aminothiophenol (4-ATP) molecules, from which a SERS enhancement factor of $3.1 \times 10^4$ was achieved.[30] This SERS enhancement factor is relatively low because the used assembly method cannot precisely control the number and allocation of Ag NPs in the Au voids. About 10% Au voids surface was coated with randomly distributed Ag NPs after the bare Au void substrate was immersed in a 20 nm Ag NPs solution. Even in the coated area, some voids may have no NP inside or may have several NPs inside (which may even aggregate to form larger NPs). Therefore, the expected CFE effect cannot be readily and efficiently realized. Furthermore, the coupling between the $^0$D mode and the NP mode in the PIC for generating the CFE requires the well control of the oblique incident angle around 60°, which is crucial for applications. Efforts were also made by other researchers on the fabrication of PICs by other methods such as the nanosphere lithography assisted by oblique deposition technique and annealing treatment.[26-30, 34-37] However, it is also difficult to obtain single gold NP dressing in each nanovoid and, moreover, the PIC array can only be arranged in closely-packed hexagonal lattice. In a word, the randomness of the NP allocation in these fabricated PIC structures is a severe problem, which seriously degrades the expected CFE effect.

Summarizing the previous studies on PICs and aiming at their practical SERS applications, we may find the following key issues to be solved. Firstly, a relatively simple PIC design is required for generating strong CFE at a SERS excitation wavelength, preferably under *normal incidence* rather than the oblique incidence at a fixed angle.[30] Secondly, small PICs with void diameter below 200 nm and NP size below 30 nm are important for generating Fano resonance and CFE in visible to near-infrared spectrum to fit SERS excitation wavelengths (as will be discussed in detail later), which however is less studied so far. Thirdly, a cost-effective and robust fabrication method is necessary to fabricate high-quality PICs with strong CFE effect. It is crucial that only one NP should be located at the bottom of each void, because the little change of the location, aggregation, and number of NPs in the void may remarkably influence the mode hybridization and thereby the induced CFE effect.[30]

In this work, we propose a strategy of realizing a high-quality plasmonic PIC nanoantenna array, by taking into account the three key issues mentioned above. It is our goal that strong CFE effect boosted by Fano resonance can be realized under normal incidence at a given wavelength. Meanwhile, a cost-effective, efficient, and reliable nanofabrication technique involving the room-temperature nanoimprinting lithography (RT-NIL)[19] and anisotropic reactive ion etching (RIE) is developed to manufacture the PICs. Since the PICs can potentially produce extremely strong CFE, which may be even stronger than those produced by some other plasmonic oligomers (such as dimers,[38] trimers [39], nanofingers[40], and nanowell, [44]nano-bricks[47]), they can be potentially exploited for single molecular sensing. To study the sensing performance of the PIC array, we employ the PIC arrays as SERS substrates for probing 4-ATP molecules and the SERS enhancement factor is evaluated experimentally.

**MATERIALS AND METHODS**

The geometry of the PIC nanoantenna unit is depicted in Fig. 1A, where a gold NP of diameter $d_p$ is located at the bottom of a truncated gold nanocavity of diameter $d_c$. The nanocavity is formed by etching a gold nanocube whose side lengths are *l* and height (which is also the height of the truncated nanocavity) is *t*. The nanoantenna array is situated on a fused silica substrate, whose lattice constants along the *x* and *z* directions are both *p*. The PIC array is assumed to be illuminated by an *x*-polarized plane wave at normal incidence from the top side. The PIC array is designed to produce strong CFE at an expected SERS excitation wavelength of 720 nm.

The optical response of the structure is numerically simulated with a commercial full-wave software CST Microwave Stuido based on the finite integral method.[41] A 3 nm thick Cr adhesion layer is considered between the gold layer and the quartz substrate. The permittivity function of Au and Cr are obtained from the experimental data and the Palik database.[42] The refractive index of the fused silica substrate is assumed as 1.5 by neglecting its small dispersion. To help understand the physics and unique properties of the PIC array, in the following we compare it with a truncated empty nanocavity (ENC) array whose structural parameters are the same as the PIC array but just without the NPs.

We have developed a reliable technique involving RT-NIL and anisotropic RIE to fabricate the PIC array, which is one of the key contributions of this work. Figure 2A describes the nanofabrication process. Firstly, two layers of photoresists, including a hydrogen silsequioxane (HSQ, Fox-14, Dow Corning Inc.) top layer of thickness 150 nm and a polymethylmethacrylate-co-polystyrene copolymer (PMMA-b-PS, a home-made imprinting resist, with PMMA/PS: 21000/64000 g.mol$^{-1}$) bottom layer of thickness 250 nm, were spincoated onto a 60 nm thick gold film coated on a fused silica substrate. After prebaking, the RT-NIL process was operated with 50 psi at room temperature. The silicon stamp is a crossed grating with cubic profiles fabricated by standard electron-beam lithography, whose lattice constants along the *x* and *z* directions are both 250 nm, and the width and depth of the cubic unit are 150 nm and 250 nm, respectively. After releasing, the HSQ in the imprinted grooves was removed by a RIE process with pure CF$_4$ (40sccm, 2Pa, 40W, 10sec). Then, another

$O_2$/$CF_4$ composite RIE process (20 sccm/ 48sccm, 2Pa, 40W, 90sec) was operated to etch the residual PMMA in the grooves and simultaneously etch the top HSQ layer.

Before the subsequent procedures, an important step of thermal annealing is necessary to induce the phase separation in the PMMA-b-PS photoresist. The sample was fast heated to 175 ℃ in one minute, kept for 5-7 min, and quickly cooled down to 25 ℃ in 10-30 sec. The PMMA-b-PS layer received an inter-surface tension between PMMA and PS during the annealing, [43] so that PMMA diffused outward while PS remained relatively stable. Consequently, the PMMA-b-PS layer become a gradient resist with varying etching resistance from center to the outer area; meanwhile, the induced tensile stress and the diffusion of PMMA caused phase separation leads to the collapse of the resist in the central area of the cube resist block, forming a resist cavity with an aggregated resist particle at the bottom of it. Then this anisotropic bilayer resist with special geometry acts as the mask for etching the underneath gold film.

With the HSQ and PMMA-b-PS bilayer mask, anisotropic RIE process was employed with $O_2$ 5 sccm/Ar 20 sccm/$Cl_2$ 10sccm plasma. During this process, the bowl-shaped cavity was transferred to the gold layer by dry etching (70W, 16Pa, 120 sec); meanwhile, since the central small resist particle in the cavity is thicker than the resist nearby, the gold underneath this resist particle was etched more slowly than the nearby area. In this way, the gold NP in the cavity was gradually formed with appropriately selected etching time and recipe. For more details on the fabrication process (including the dry etching of the gold film, the preparation of the bilayer resist mask, the formation of the gold NP during RIE, the control of the gold NP size and position, and the stability and purity of the fabricated gold PIC array), readers may refer to Section 1 of the Supporting Information.

Note that the size of the gold NPs can be finely controlled by tuning the anisotropic RIE parameters. For example, with a RIE recipe of $O_2$ 10 sccm/Ar 20 sccm/$Cl_2$ 10 sccm, 16Pa, 70W, and 150 sec, the ENC array without gold NPs can be formed. After etching, the residual resists were completely cleaned by soaking the sample in an acetone solvent for 30 min and implementing an ultrasonic bathing for 3min. The cleaned PIC nanostructures are pure gold without any residual resist, which was verified by energy dispersive X-ray spectroscopy characterization (see the Section 1 of the Supporting Information for more details.) The quality and geometry of the fabricated structures were examined by a scanning electronic microscope (SEM, FEI Serion 2000).

**RESULTS AND DISCUSSION**

The PIC array is designed to produce strong CFE at an expected SERS excitation wavelength of 720 nm. As mentioned above, strong Fano resonance may be generated in the PIC array via the coupling between the NP mode and a cavity mode of the nanovoid. Therefore, the array should be designed so that the concerned cavity resonance is around the target wavelength 720 nm. By numerical simulations (see details in Section 2 of the Supporting Information) and according to previous studies, [26-30] we found that although the cavity resonance modes are generated by individual cavities, their spectral positions can be effectively influenced by the lattice constant when the cavities are arranged in a periodic lattice. Therefore, by numerical optimization and also by taking into account the fabrication capabilities, we chose the array period $p = 250$ nm, the nanocube width $l = 150$ nm (by which the coupling between the adjacent PIC units is avoided), the cavity diameter $d_c = 120$ nm, and the height of the truncated cavity $t = 60$ nm. With these parameters, the far-field reflectance and transmittance spectra of the ENC array were simulated and shown in Fig. 1C. It is seen that the reflectance spectrum has two evident resonance peaks in the wavelength range of 700 nm to 1000 nm, which are attributed to the $^1D_+$ and $^1P_+$ modes of the truncated nanocavity under normal incidence.[28] This can be verified by the simulated field distributions in the ENC in Figs. 3E and 3F. Note that our ENC array is situated on a silica substrate, which is different from the case of Ref. [28] where the empty nanovoids are embedded in an infinitely thick gold substrate. This explains why the simulated field distributions of the two modes are slightly different from those in Ref. [28]. But the key features of the mode profiles remain the same, indicating the excitation of the two modes. With these selected void parameters, the $^1D_+$ mode is tuned to be very close to the target wavelength 720 nm. We are especially interested in this mode because the field hot spots of this mode (see Fig. 3E) are mainly located at the bottom of the void so that it may have strong coupling with the field of the gold NP.

Then the last important parameter to be determined for the PIC array is the size of the gold NP. As seen from the simulated reflectance spectrum of a PIC array with 20 nm gold NPs (i.e., the diameter $d_p$= 20 nm) in Fig. 1B, the main consequence of introducing a gold NP in each nanovoid is the generation of an additional peak between the $^1D_+$ and $^1P_+$ peaks, although the peak positions of the $^1D_+$ and $^1P_+$ cavity modes also shift slightly due to the perturbation by the gold NP. This additional peak is caused by a so-called "NP mode" that is induced by the coupling between the gold NP and its image.[30] Compared with the nanocavity modes $^1D_+$ and $^1P_+$ that cause relatively broad resonances (which are super-radiant bright modes), the NP mode is a sub-radiant dark mode and causes narrow resonance. When the NP mode is close to the $^1D_+$ mode, they have strong interference leading to strong Fano resonance, manifesting as an asymmetric resonance peak in the transmittance spectrum of the PIC array (see the red curve in Fig. 1B), which does not exists in the ENC array (see the red curve in Fig. 1C). We have also thoroughly studied the influence of the NP size on the Fano resonance peak position as well as the maximum field enhancement factor at the target wavelength 720 nm, as seen from Figs. 1E and 1F. Clearly, with the increase of the NP size, the Fano resonance peak has a red shift (as indicated by the arrows in Fig. 1F). When $d_p$= 20 nm, the Fano peak is tuned to be exactly at the target wavelength 720 nm and the corresponding field enhancement factor also reaches the maximum at this wavelength (as seen from Fig. 1E). So we select $d_p$= 20 nm in our PIC array design. Note that when the NP size is further increased, the NP mode peak may further red shift so that it has weaker interference with the $^1D_+$ mode but stronger interference with the $^1P_+$ mode (as seen from the pink curve in Fig. 1F).

To verify the far-field responses of the PIC array, we have measured the transmittance of the fabricated PIC and ENC arrays by a spectroscopic ellipsometer VASE (J. A. Woollam Co.) in the wavelength range from 400 nm to 900 nm under normal incidence, as shown in Fig. 1D. Clearly, the experimental spectra agree well with the theoretical predictions, verifying the expected Fano resonance. The slight differences of the spectra are probably caused by the deviation of the structural parameters of the fabricated samples from theoretical designs.

As discussed above, the ultimate goal of generating and tuning Fano resonance by mode hybridization in the plasmonic PIC system is to produce strong CFE effect. To explore this, we have simulated field distributions in the PIC array in Figs. 3A-3D. It is seen that the $^1D_+$ mode (Fig. 3A) and $^1P_+$ mode (Fig. 3D) still keep their mode profile features (although perturbed somehow by the gold NP). The Fano interference of the $^1D_+$ mode with the NP mode (Fig. 3B) leads to remarkable CFE at the target wavelength of 720 nm (as seen from Fig. 3C), leading to not only the greatly enhanced field hot spots but also the slight change of distribution of the hot spots in the cavity. This is very important and beneficial for enhancing the light-matter interaction in the cavities. The simulated field enhancement factor (i.e., the normalized field amplitude $|\mathbf{E}/\mathbf{E}_0|$ where $\mathbf{E}_0$ is the amplitude of the incident electric field) is as high as 500, while in contrast the enhancement factor in the ENC array without the CFE effect is only 150. Thus the CFE effect can significantly improve the field enhancement factor.

It should be noted that in all the above simulations we did not take into account the finite contact area between the gold NP and the gold void. That is, the contact between the NP and the void was assumed to be an ideal point. Actually, in the fabricated samples, such a physical contact formed during the etching process must have non-zero area. In Section 3 of the Supporting Information, we have studied the influence of the contact area on both the far-field and near-field responses of the PIC array. It is shown that with the increase of the contact area, the Fano resonance peak may have a slight decrease (although the resonance peak profile is almost unchanged) and the near-field enhancement may also decrease correspondingly. For example, when the radius of the contact area is assumed to be 4 nm, the maximum near-field enhancement factor in the PIC would decrease to about $4\times10^2$. This means that the calculated field enhancement factor of $1.2\times10^3$ above is the most optimistic one. In practical samples, the field enhancement may be degraded due to the finite contact area as well as other fabrication defects.

To corroborate the strong field enhancement in the PIC array, we have fabricated some samples of ENC and PIC arrays, as seen from the SEM images in Figs. 2 B and 2C, respectively. The insets show the blowup of the top-view of the samples. From Fig. 2C, it is seen that only one gold NP is situated in each nanocavity, verifying the reliability and quality of the fabrication method. While in the ENC array in Fig. 2B, there is no gold NP in the nanocavity.

As a demonstration of the SERS performance of the PIC array, we have conducted a comparative experiment by using the PIC array and ENC array as SERS substrates for probing 4-ATP molecules (Sigma-Aldrich Co., with bulk concentration $0.1\mu M$). The reference Raman signal was characterized from an unpatterned area of the gold film. All the Raman spectra removed the baseline and were fitted using a Guassian-Lorentzian line shape. In the SERS experiment, 4-ATP molecules were dosed onto the sample substrates via the chemical bonds between gold and sulfur atoms on the surface when the substrates were soaked for 3 hours in the 4-ATP solution, and washed several times with ethanol to obtain approximate a monolayer of 4-ATP molecules on the samples.

The back-scattering Raman spectra were taken and collected by a LabRam microRaman system (Jobin-Yvon/ISA HR). A super-continuum light source (Fianium, SC-400-PP) was tuned to emit light at the excitation wavelength of 720 nm with an output power 0.6mW and integration time 10 s. A thermoelectrically cooled charge-coupled device (CCD) array detector and an BH-2 microscope with $100\times$short-focal-length objective lens with a 0.5 aperture were employed. The Raman signal probing area was about 1μm in diameter. The SERS spectra were collected in the Raman shift range of 400 cm$^{-1}$ to 1800 cm$^{-1}$.

Figure 4 shows the measured SERS spectra of the 4-ATP molecules on the PIC array, the ENC array, and the reference gold film. Note that the reference spectrum is multiplied by a factor of 10. It is seen that the SERS signal collected from the PIC array is the strongest, which is about four times stronger than the spectrum collected from the ENC array. The SERS enhancement factor (*EF*) can be estimated by: [6]

$$EF = \frac{I_{SERS}/N_{surf}}{I_{bulk}/N_{bulk}} \quad (1)$$

where $I_{SERS}$ and $I_{bulk}$ are the Raman signal intensities at the characteristic vibration mode at 1078 cm$^{-1}$ of 4-ATP molecules doped on the nanostructured samples and on the surface of the reference flat gold film, respectively; $N_{surf}$ and $N_{bulk}$ are the estimated numbers of 4-ATP molecules doped on the sample and bulk 4-ATP molecules in the probing light spot area, respectively. The details on the estimation of $N_{bulk}$ and $N_{surf}$ are given in Section 4 of the Supporting Information. Thus, according to Eq. (1), the experimental SERS enhancement factor of the PIC array can be estimated to be 2×10$^7$. This is a relatively conservative estimate. Since the collected Raman signal from the PIC array mainly comes from the gold NP with strong field hot spots, the real number of 4-ATP molecules $N_{surf}$ adsorbed on the NP would be much smaller. If this is taken into account, the estimated experimental SERS enhancement factor can be even higher (up to 10$^9$).

The SERS experiment provide persuasive evidence that the PIC array has greatly improved SERS performance, due to the CFE effect boosted by Fano resonance in the PIC nanoantennas. Therefore, the PIC array has great potential to be used as highly active SERS substrate.

**Conclusions**

We have proposed a plasmonic PIC array consisting of nanoantennas in each of which a 20 nm gold NP is placed at the bottom of a 120 nm truncated hemispherical gold nanovoid. Through plasmonic mode hybridization in this multiscale system, strong Fano resonance can be generated due to the simultaneous excitation and inter-coupling of the NP mode and a high order cavity mode, which then significantly boost the cascaded field enhancement in the PIC antenna. The enhancement factor of the localized field amplitude can be up to $10^2$-$10^3$, which is one order of magnitude higher than that in a reference ENC array. To realize the PIC array with such high performance, we have developed a cost-effective, efficient and reliable nanofabrication method involving RT-NIL and anisotropic RIE. With this technique, it is assured that only one gold NP is located in each nanovoid and the NP size can be finely controlled to be around 20 nm, which are crucial for realizing the significant CFE effect in the PIC array. As a demonstration of the CFE performance of the PIC array, the PIC samples were fabricated and employed as SERS substrates for detecting 4-ATP molecules. The achieved experimental SERS *EF* is as high as $2 \times 10^7$, demonstrating the strong field enhancement in the PIC array and its great potential of applications.

## ACKNOWLEDGMENTS

We acknowledge the support by the National Natural Science Foundation of China (Projects No. 11474180, and No. 61227014) and the Ministry of Science and Technology of China (Project No. 2011BAK15B03).

**Figure Captions**

**Figure 1.** A) Schematic geometry of the unit cell of a two-dimensionally periodic PIC array, which is illuminated by an *x*-polarized plane wave under normal incidence. B) and C) are the numerically simulated far-field transmittance and reflectance spectra of the PIC array and a reference ENC array, respectively. The resonance peaks of the $^1D_+$ and $^1P_+$ cavity modes, the NP mode, and the Fano resonance are indicated by arrows. D) The experimentally measured transmittance spectra of a fabricated PIC array and an ENC array. E) The calculated field amplitude enhancement factor in the PIC array with respect to the increase of the gold NP diameter. F) The simulated transmittance spectra of the PIC array with different gold NP sizes, where the Fano resonance positions are indicated by arrows.

**Figure 2.** Schematic of the fabrication process of the PIC array by using a technique based on RT-NIL and anisotropic RIE. B) SEM image of a fabricated ENC array, in which the inset shows the blowup of the sample. (C) SEM image of a fabricated PIC array, in which the inset shows the top-view blowup of the sample.

**Figure 3.** Simulated normalized electric field amplitude distributions in the PIC array and in the ENC array. A) The $^1D_+$ mode in the PIC array at wavelength 697 nm. B) The NP mode in the PIC array at wavelength 743 nm. (C) The Fano resonance in the PIC array at wavelength 720 nm. D) The $^1P_+$ mode in the PIC array at wavelength 900 nm. (E) The $^1D_+$ mode in the ENC array at wavelength 724 nm. (F) The $^1P_+$ mode in the ENC array at wavelength 915 nm.

**Figure 4.** Experimentally measured Raman spectra of the 4-ATP molecules on different substrates: a PIC array, an ENC array, and an unpatterned gold film for reference. The 1078 cm$^{-1}$ band is the fingerprint vibration wavenumber of the 4-ATP molecules.

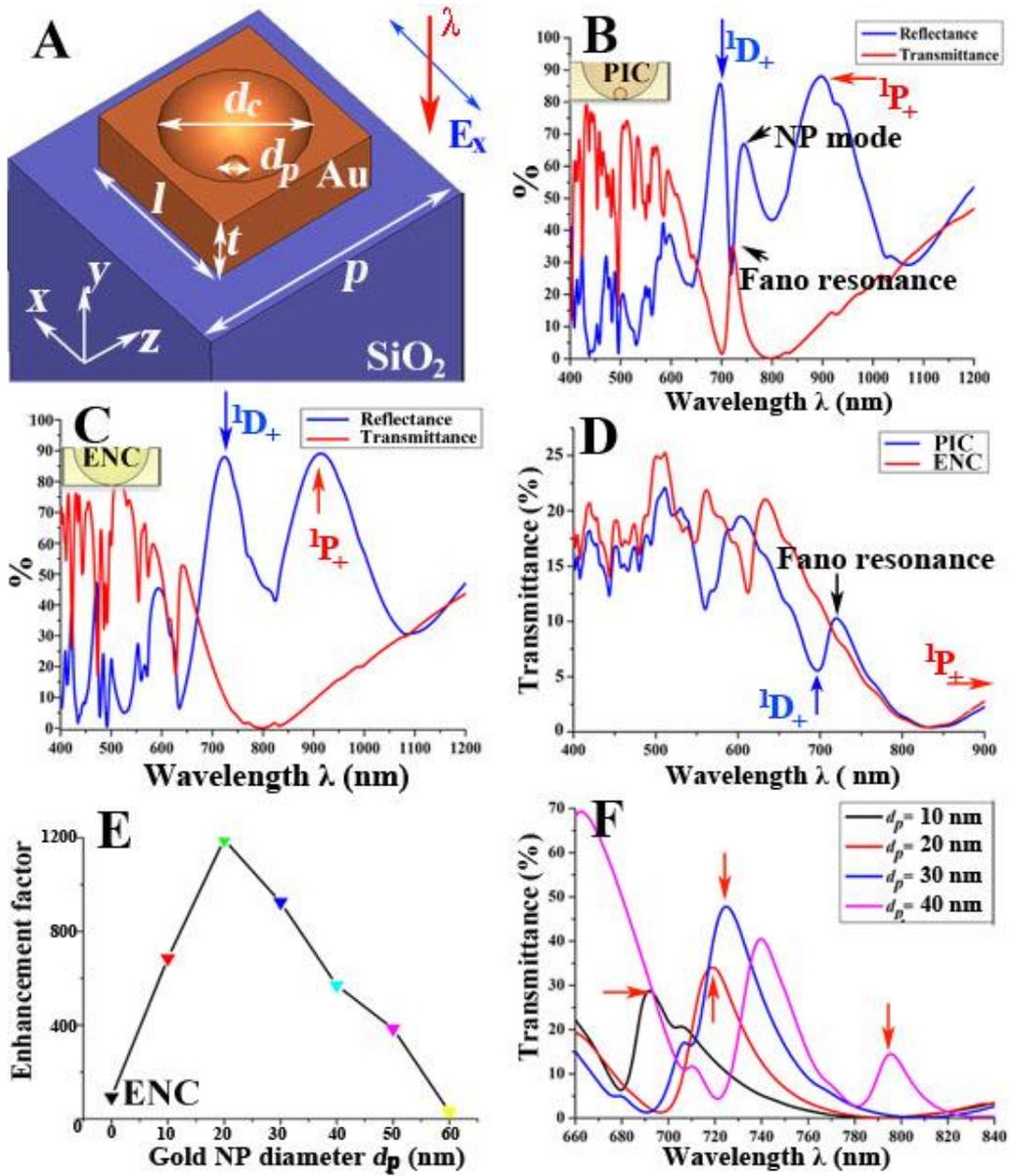

**Figure 1**

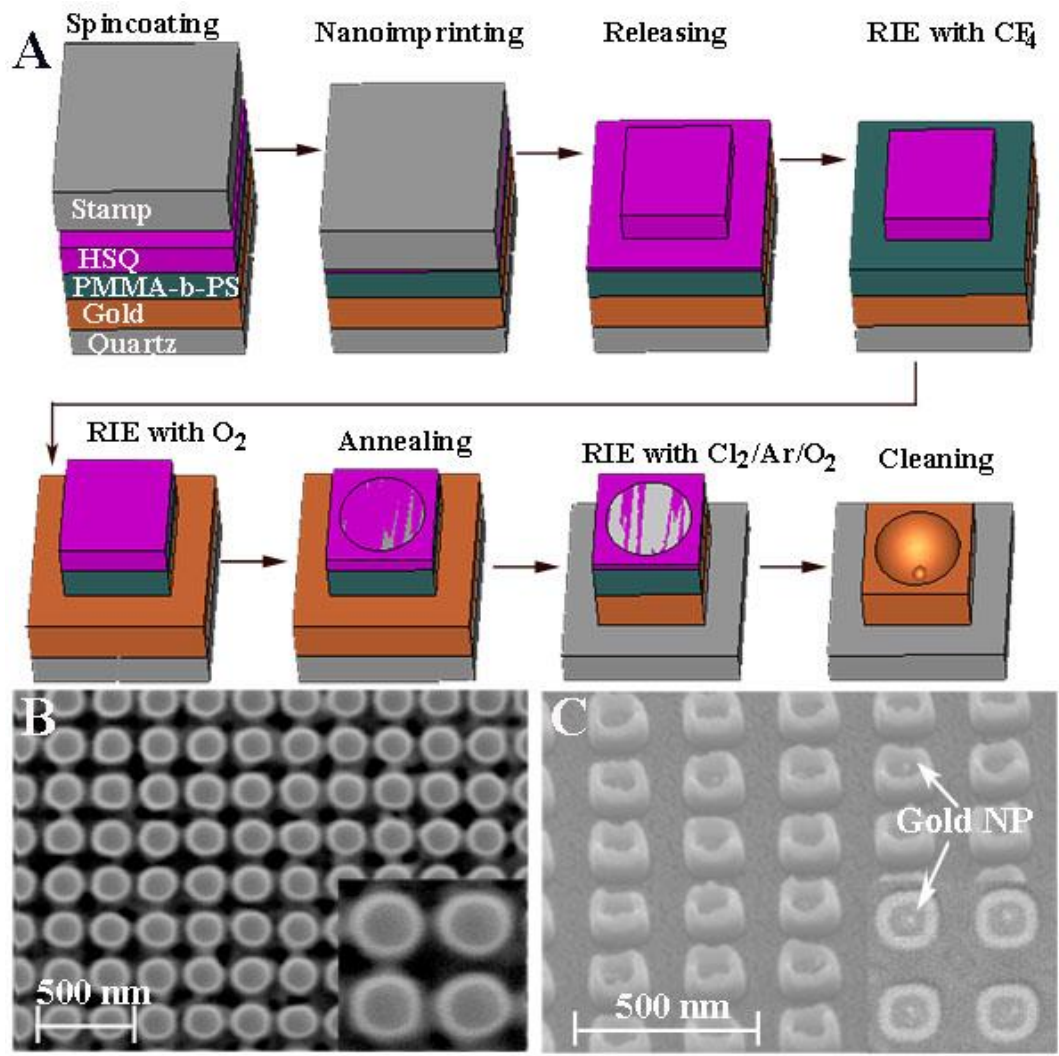

**Figure 2**

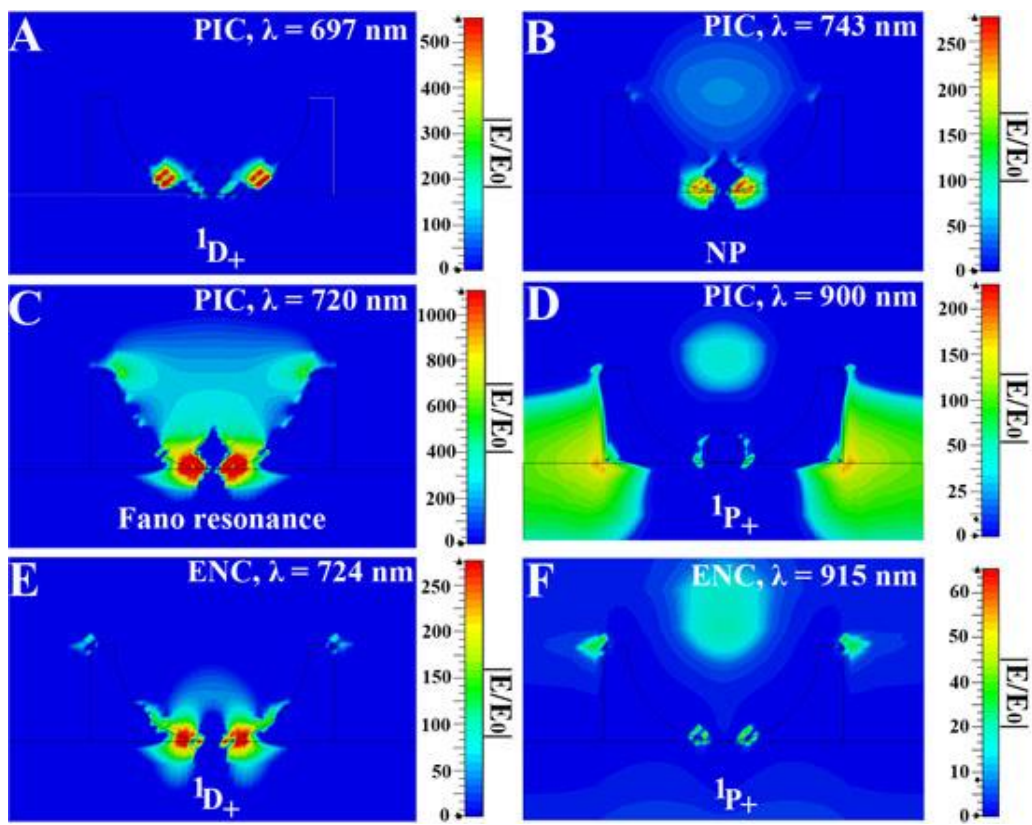

**Figure 3**

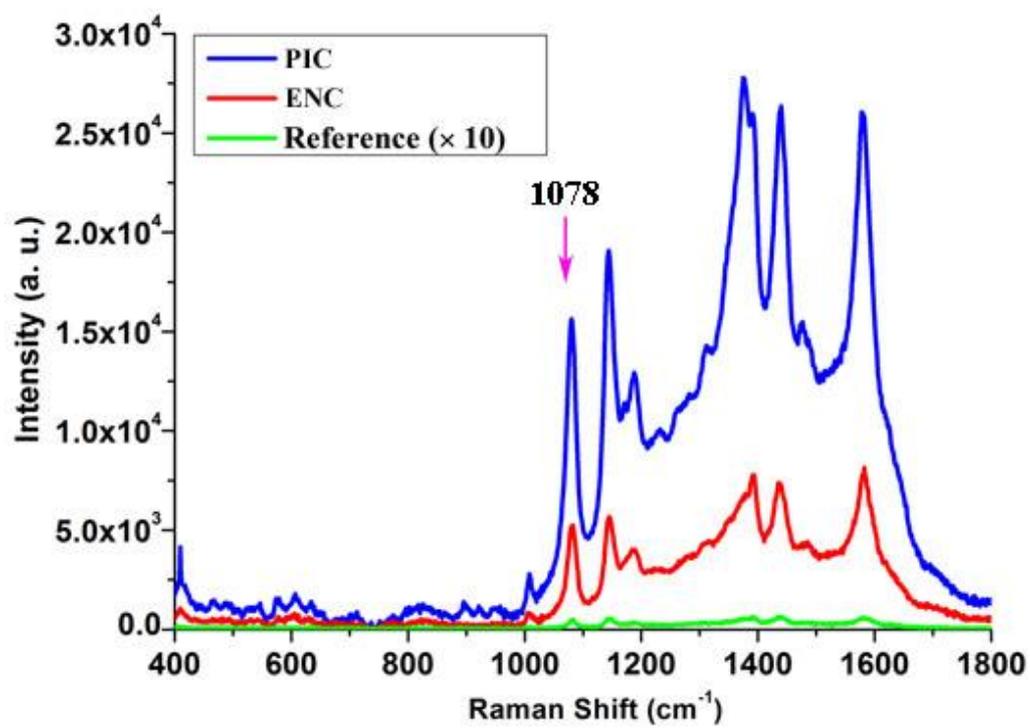

**Figure 4**